# Formalizing typical crosscutting concerns

Marius Marin

# Formalizing typical crosscutting concerns

Marius Marin
*Software Evolution Research Lab*
*Delft University of Technology*
*The Netherlands*
A.M.Marin@ewi.tudelft.nl

## ABSTRACT

We present a consistent system for referring crosscutting functionality, relating crosscutting concerns to specific implementation idioms, and formalizing their underlying relations through queries. The system is based on generic crosscutting concerns that we organize and describe in a catalog.

We have designed and implemented a tool support for querying source code for instances of the proposed generic concerns and organizing them in composite concern models. The composite concern model adds a new dimension to the dominant decomposition of the system for describing and making explicit source code relations specific to crosscutting concerns implementations.

We use the proposed approach to describe crosscutting concerns in design patterns and apply the tool to an open-source system (JHotDraw).

## 1. INTRODUCTION

Aspect-oriented programming languages provide mechanisms to enable modularization of crosscutting concerns, such as *pointcut and advice* or *introductions* [14, 7, 13]. These mechanisms address general symptoms of crosscuttingness, like scattering (i.e., a concern lacks localization being spread over several modules) and tangling (i.e., a module implements a core concern, but also crosscutting aspects).

These symptoms and language mechanisms alone, however, do not ensure consistency for defining and describing crosscutting concerns. To understand concerns specific to logging mechanisms, design patterns, or implementation of business rules, one has to rely on unconstrained descriptions or sample code examples [10, 16, 20, 1]. Moreover, similar concerns are sometimes described by different names (e.g., *contract enforcement* [1] and *policy enforcement* [16]).

The absence of a consistent and coherent system for defining and referring crosscutting concerns hinders their comprehension and identification in existing code (a.k.a. aspect mining). The wide variety of examples describing crosscutting functionality and proposed aspect solutions does not distinguish common properties to group similar concerns and to separate between different groups. Hence, it is difficult to define requirements for aspect mining techniques, to describe and compare their results [17, 2] without an answer to the research question:

*What are the crosscutting concerns that we try to identify and how to consistently describe and reason about these concerns*?

To address this question we have previously outlined a classification system for crosscutting concerns based on *sorts*, and proposed an informal set of sorts [18, 19]. Crosscutting concern sorts are *generic* and *atomic* concerns described by their specific symptoms (i.e., implementation idiom in a non-aspect-oriented language), and a (desired) aspect mechanism to modularize a sort's instances with an aspect-based solution.

This work provides a formalized description of the crosscutting concern sorts and turns the list of sorts into a catalog. The catalog describes each sort as a relation between code elements and provides query templates to formalize these relations. The catalog is built on an in-depth analysis of crosscutting concerns reported in literature and practical experience with source code exploration and analysis for identification of crosscutting concerns in software systems of considerable size like PetStore, JHotDraw and Tomcat [17].

The contributions of the paper lie in several directions:

**A catalog of generic crosscutting concerns** to ensure consistency and coherence in addressing crosscutting functionality. The paper describes unified criteria for grouping crosscutting concern as relations between code elements and provides detailed examples for each entry in the catalog.

**A query component and templates for each sort** to formalize the relation underlined by the concerns of the sort. The templates support the development of query libraries for crosscutting concerns. Moreover, the formalization of the queries identifies relations relevant for describing crosscutting concerns and allows for extensions of existing query languages for capturing such concerns.

**Crosscutting concerns documented** using sort queries for the catalog of design patterns described by Gamma et al. [8] and for a popular open-source system of relevant size. This documentation is available for further references as a common benchmark for aspect mining [2]. The paper also discusses in significant extent a number of relevant concerns in this system and how sorts help in describing and documenting them.

**Advanced tool support** for querying sort instances in code under analysis and building concern models to describe crosscutting relations in software systems. A concern model allows for grouping sort instances described by queries into composite concerns. This model adds a new dimension to the traditional representation based on a dominant decomposition of an object-oriented system [25]. This dimension relates relevant program elements in the context of a concern implementation and makes this relation explicit.

The organization of relevant relations describing generic crosscutting concerns and the proposed formalization using queries is aimed at supporting crosscutting concern comprehension. Nevertheless, the descriptions of concerns using sorts contribute a first step towards refactoring to aspect-oriented programming by documenting typical crosscutting relations. Aspect-oriented languages could use the contributions of this paper to further support the refactoring of concern sorts to aspect solutions.

The rest of the paper is organized as follows: The next section presents related work on describing crosscutting concerns and how the existing approaches compare to the one proposed in this paper. In Section 3 we introduce the classification of crosscutting concerns on sorts, followed in Section 4 by the query component and the template to describe the sorts. Then, we present a catalog of crosscutting concerns sorts together with examples of relevant concerns and template queries for each sort.

In Section 5 we look at how the proposed sorts cover the crosscutting elements occurring in the implementation of design patterns. Then, in Sections 6-7, we present the tool support and the results of applying the tool to an open-source system for describing instances of crosscutting concerns and organizing them in a composite model. We conclude with lessons learnt, opportunities for improvements and directions for future work.

# 2. RELATED WORK

## 2.1 Support for defining and describing crosscutting concerns

The Feature Exploration and Analysis Tool (FEAT) and the Concern Manipulation Environment (CME) provide tool support for grouping program elements related by a crosscutting concern implementation and organizing them into more complex, composite structures.

FEAT organizes program elements that implement a concern in *concern graphs* [22]. The user can add elements to a concern graph by investigating the incoming and outgoing relations to and from an element that is part of the concern implementation. The elements in a concern graph are classes, methods or fields connected by a *call, read, write, check, create, declare,* or *superclass* relation.

Although the tool allows to add relations to the graph describing a concern, the focus is on the elements participating in the implementation of the concern. The navigation for understanding a concern and incrementally building its graph representation is from a root (class) element to other elements in the relation chain. That is, a concern is described by its elements, and an element is described by its relations. Unlike FEAT, the sorts-based approach that we propose uses relations as the main representation of a concern and builds concern models based on these relations.

CME allows for more complex queries than FEAT to define relationships between code elements, especially by providing options for restricting the query domain [11]. The query and its definition can be saved as an element in the CME concern model to describe a concern intensionally. The output of the query can also be saved to describe a concern extensionally.

However, neither CME nor FEAT provide a coherent, standard way to document crosscutting functionality: no consistency is imposed for documenting instances of the same or similar concerns and no distinction is made between different concerns documented by the same or similar queries. This prevents uniformity in addressing and describing crosscutting functionality and reflects upon the efficiency of documenting concerns with, for instance, CME queries: what is the right query for describing a particular crosscutting concern, like, for example, a simple pre-condition check? Is this query the same as the other queries documenting pre-condition checks in the same system? Would it be possible (and desirable) to distinguish between a standard query documenting calls for pre-condition checks and a query to document logging calls? What queries are needed to consistently document complex relations like a design pattern implementation?

## 2.2 Refactoring to aspects

The work on refactoring to aspect-oriented programming shows a significant number of examples of crosscutting concerns.

The present approaches to aspect refactoring can generally be distinguished by their granularity. The group of aspect refactorings proposed by Laddad covers a significant number of situations where crosscutting functionality could occur in an application [16, 15]. Some of these refactorings are low level ones, closely associated with symptoms of a large variety of concern implementations, such as *extract method calls into aspects* or *extract interface implementation*. Yet most of the refactorings are relatively complex and include design patterns, transactions management, or business rules. These complex refactorings generally involve more than one concern to be refactored. Transaction management, for instance, implies to take care of committing or rolling back a transaction for a given operation, to ensure that the same connection is used for all the updates, or that there are no undesired calls on connection objects.

In the literature describing this group of refactorings, there is no categorization of the various refactorings proposed nor a specific classification of the concerns involved [1].

Hannemann et al. [9, 10] propose role-based refactorings for object-oriented design patterns [8]. The refactoring relies on a library of abstract descriptions of the patterns and their role elements, and instructions to refactor to an aspect solution after mapping code elements to the abstract roles. This approach is a step further towards defining generic, abstract solutions to typical problems that involve crosscutting functionality, such as applying design patterns. However, design patterns implementations only define the context into which the crosscutting occurs, they are not concerns. Moreover, the design patterns typically exhibit multiple crosscutting concerns, some of them sharing common properties, as we shall see in Section 3.1.

[1] Work in [16, 1] mentions an aspect classification based on the phase of the software lifecycle at which the aspects occur: development and production aspects. This classification is not related to the discussion proposed here.

Finer-grained refactorings have been proposed in form of code transformations catalogs [20] and AspectJ laws [3]. These transformations can occur as steps in the aspect refactoring of a crosscutting concern, but they are oblivious to the refactored concern. They describe mechanics of migrating Java specific units to AspectJ ones (e.g., *Extract Fragment into Advice, Move Method/Field from Class to Inter-type*).

## 2.3 Query languages and tools

From the discussed tools and approaches to describing and organizing crosscutting concerns, only CME provides its own query language: Panther is a pattern-matching language that allows to search for program units and query relationships between them [24]. This language, however, lacks a precisely defined syntax, as well as a complete implementation.

Alike CME, JQuery is a code browser developed as an Eclipse plug-in. JQuery uses a logic query language (TyRuBa) similar to Prolog [4]. The TyRuBa predicates supported by JQuery cover all the relationships defined by FEAT and include many others. For example, JQuery supports a number of predicates for checking the type of an argument; the recognized types are compilation units, class members, errors, tasks, warnings, etc. Among the supported binary relations some are not present neither in FEAT nor in Panther, such as relations for thrown exceptions.

Despite being more flexible than CME, JQuery provides a less friendly query syntax. For example, a simple query for returning the classes implementing a specified interface can be written in Panther like:

```
sourceof(relationship
    implements(type *, interface FigureChangeListener)); .
```

The same query in JQuery is more difficult to write and understand:

```
interface(?I), name(?I,FigureChangeListener),
                                    implements(?C, ?I); .
```

Sextant is a tool similar to JQuery, which also enables to query different kinds of artifacts [5]. The tool stores an XML representation of a software's artifacts and uses the XQuery [2] language to define queries across these artifacts. However, this is less relevant to our approach, which is aimed at describing crosscutting relations in source code.

Eclipse IDE provides an advanced programming interface for querying source code relations and various views for visualization of results. This support is more mature than the one in CME and provides a significantly better performance (speed and resources) than JQuery. Moreover, both CME and JQuery build their own internal representations of the code to be analyzed, parallel with the one available through Eclipse. This implies recompilation of the code to be analyzed by the two tools, and additional use of resources.

From the available options, we have chosen to use a pseudo query language that resembles CME's Panther language for formalizing the sort queries. This provides us with a more elegant syntax to ease understanding. However, we have implemented our tool support as a plug-in that uses Eclipse's infrastructure for querying source code and reporting results. This solution provides us with improved performance.

[2] www.w3.org/TR/xquery

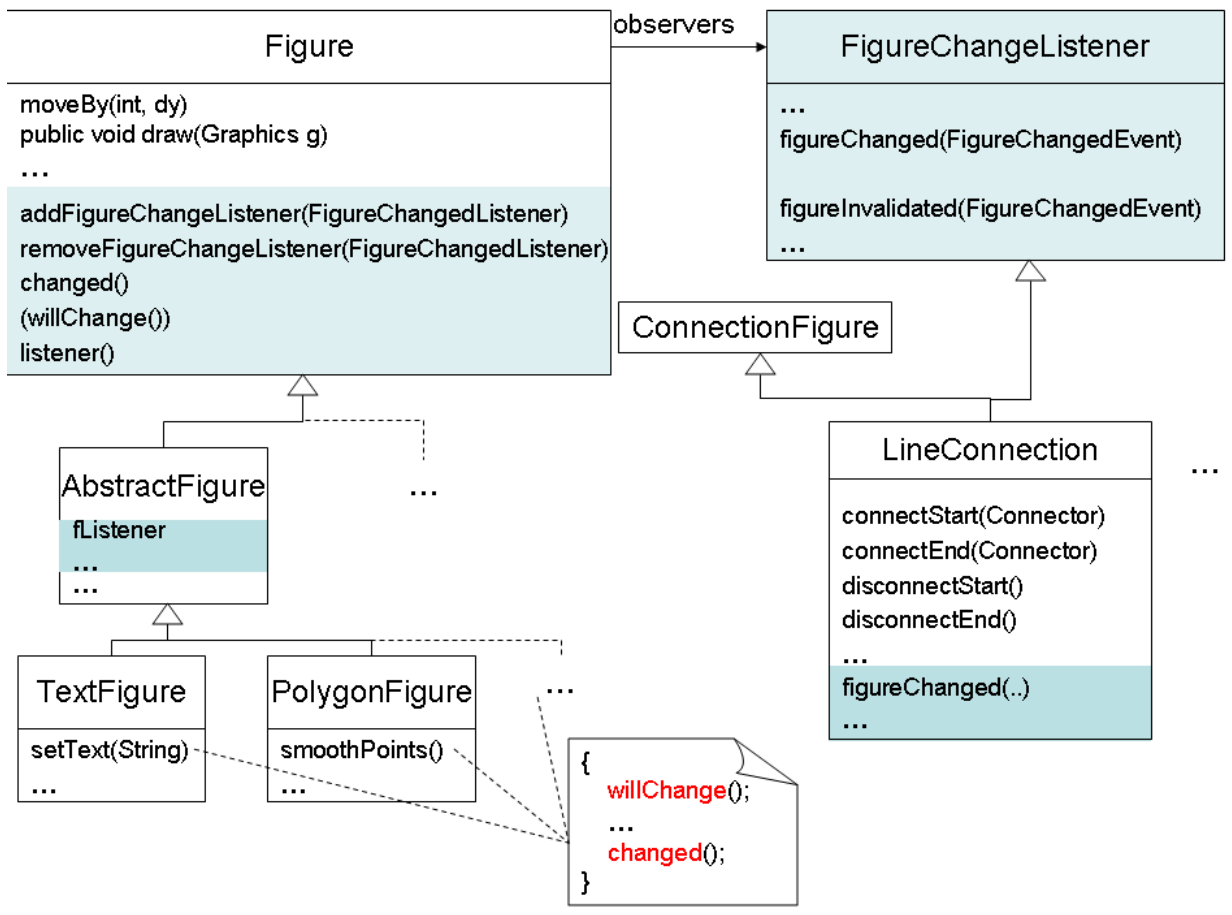

**Figure 1: FigureChanged Observer**

# 3. CROSSCUTTING CONCERN SORTS

The (dominant) decomposition of a software system provided by object-oriented languages permits to make certain relations explicit in source code. Such relations include inheritance, call relationst, encapsulation of properties and containment of actions, etc. Not all relations between program elements, however, can be made explicit by object-oriented languages: the relation between a set of methods consistently invoking the same action as part of a common requirement is hindered in the code and possibly made explicit by comments and documentation. Similarly, classes that implement multiple roles (defined by different interfaces) relate secondary roles to a primary one in a given context. We refer to such relations "hidden" in the code as crosscutting concerns and further look at how these can be categorized for a consistent reference. The aim of this work is to provide a coherent system for referring crosscutting functionality, to relate crosscutting concerns to specific implementation idioms and to formalize their underlying relations.

## 3.1 Motivating example

Design patterns, and most notably the Observer pattern, are typical examples of problems exhibiting crosscutting functionality [10]. A standard implementation of the Observer pattern introduces specific relations through the participant-roles and the collaborations described by Gamma et al. [8]. The classes that participate in an implementation of the pattern play either the role of the Observer or that of the Subject. The two roles define specific members, typically abstracted in an interface definition: The Subject role knows its observers and defines methods to allow attaching and detaching Observer objects, as well as a notification method to inform the observers of changes in the Subject's state. Actions that change the state of the Subject consistently invoke the notification method. The Observer defines a specific method to receive notifications from the Subject and update its own state consistently.

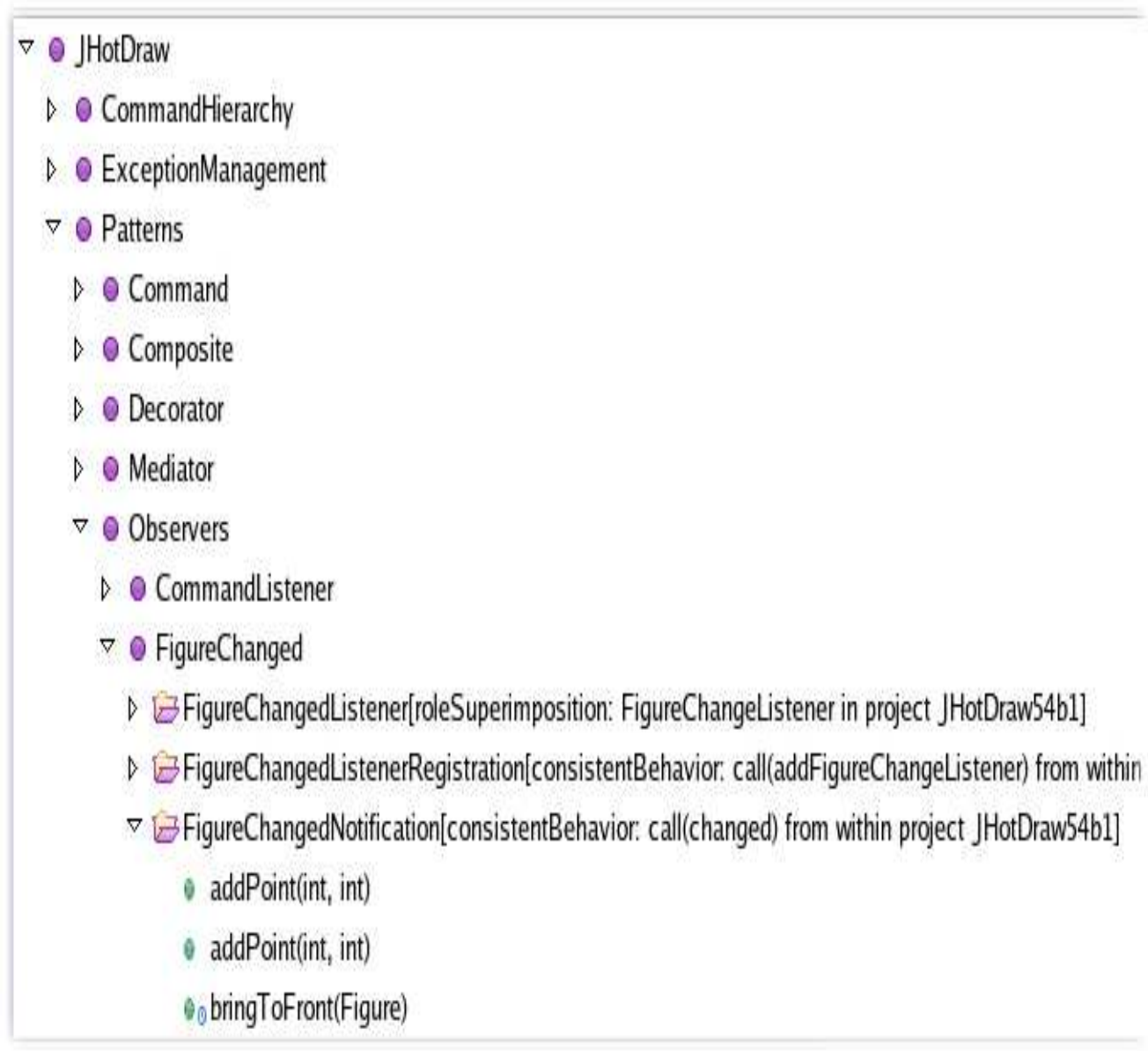

**Figure 2: Concern model for FigureChanged Observer instance**

Figure 1 shows an instance of the Observer pattern in a drawing application. The *Figure* elements play the Subject role and declare role-specific members like the `changed()` method to notify observers, the methods for attaching and detaching observer objects, and fields in concrete classes to store the references to the observers. The *FigureChangeListener* interface defines the Observer role.

The atomic crosscutting relations that occur due to the implementation of the Observer pattern, like the consistent notification of observers or the implementation of multiple roles by the classes participant in the pattern, are not, however, specific to this pattern only. Implementation of multiple roles occurs in other patterns as well, like the Visitor, or in mechanisms for implementing persistence, like Java's *Serializable* mechanism. Such relations exhibit a common implementation idiom, namely implementation of multiple interfaces or direct implementation of members that can be abstracted in an interface definition.

Similarly, the idiom for the consistent notification consists of scattered method calls to a common functionality. This idiom is shared by other known crosscutting concerns, like, for example, consistent tracing operations.

The common relations and idioms specific to crosscutting concerns like consistent notification or tracing suggest that we can define generic crosscutting concerns, which describe common properties. We call these generic crosscutting concerns, *sorts*. The two sorts in an implementation of the Observer pattern are *Role superimposition* and *Consistent behavior*. The Role superimposition, for example, has two instances in the Observer pattern, specific to the two roles for the participants in the pattern.

## 3.2 Starting point

This section summarizes our past contribution to defining sorts of crosscutting concerns.

A crosscutting concern sort is a generic description of a class of concerns that share a number of properties:

- an intent (behavioral, design or policy requirements),
- a specific implementation idiom in an (object-oriented) language and
- a (desired) aspect language mechanism that supports the modularization of the sort's concrete instances.

Examples of aspect mechanisms include *pointcut and advice* or *introduction*, as in AspectJ [13, 1].

Concrete implementations of a crosscutting concern sort in source code represent *sort instances*.

Together with defining the sorts, we have also provided an informal list of canonical sorts with a focus on their refactoring to aspect-oriented programming [18].

In the present paper we address the sorts as *un-modularized generic relations* between program elements. To capture and make explicit these relations, we provide a query component for describing sorts.

We revise the previously proposed list of sorts and provide a formalized description of the sorts. Furthermore, we organize the sorts in a catalog that describes each sort in detail.

Two of the sorts in the previously proposed list have been merged as they differed only by the refactoring solutions. Soares et al. [23] proposed a mechanism for introduction of *throws* clauses for addressing the Exception propagation sort described in Section 4.4. This is very much an alternative to the AspectJ approach that uses exception softening. The distinction between the various refactoring approaches is less important to our current focus on describing concerns by their defining relations.

## 3.3 A concern model based on crosscutting concern sorts

Crosscutting concern sorts are, by definition, atomic elements. The atomicity of a sort provides a consistent granularity level for classifying crosscutting concerns by common properties, but also gives the level of complexity for the relations to be expressed by a sort instance. To relate sort instances in more complex relations, like participation in design patterns implementation, we use a composite *concern model*. A concern model organizes concerns in a hierarchical structure. Relations describing sort instances are always leaf elements and children of a composite concern model. Each model can be a child of a super-model. A concern model for a complex relation $R$ can be formalized as:

$$ConcernModel(R) = \bigcup_{i=1}^{n} SI_i \cup \bigcup_{j=1}^{m} SR_j$$

This model describes a relation R as a composition of sort instances $SI_i$ and composite sub-relations $SR_j$. Figure 2 shows the concern model for the Observer implementation previously discussed. The composite *FigureChanged* model groups instances of sorts like Role superimposition and Consistent behavior that participate in the implementation of the Observer pattern for figure changes. A sort instance is described by a a given name and an associated query together with the results of this query. A concern model is also described by a given name.

The FigureChanged relation is part of parent, custom-defined relations, like the one grouping all the instances of the Observer pattern in the JHotDraw project. In this case, the project corresponds to the top-level concern model.

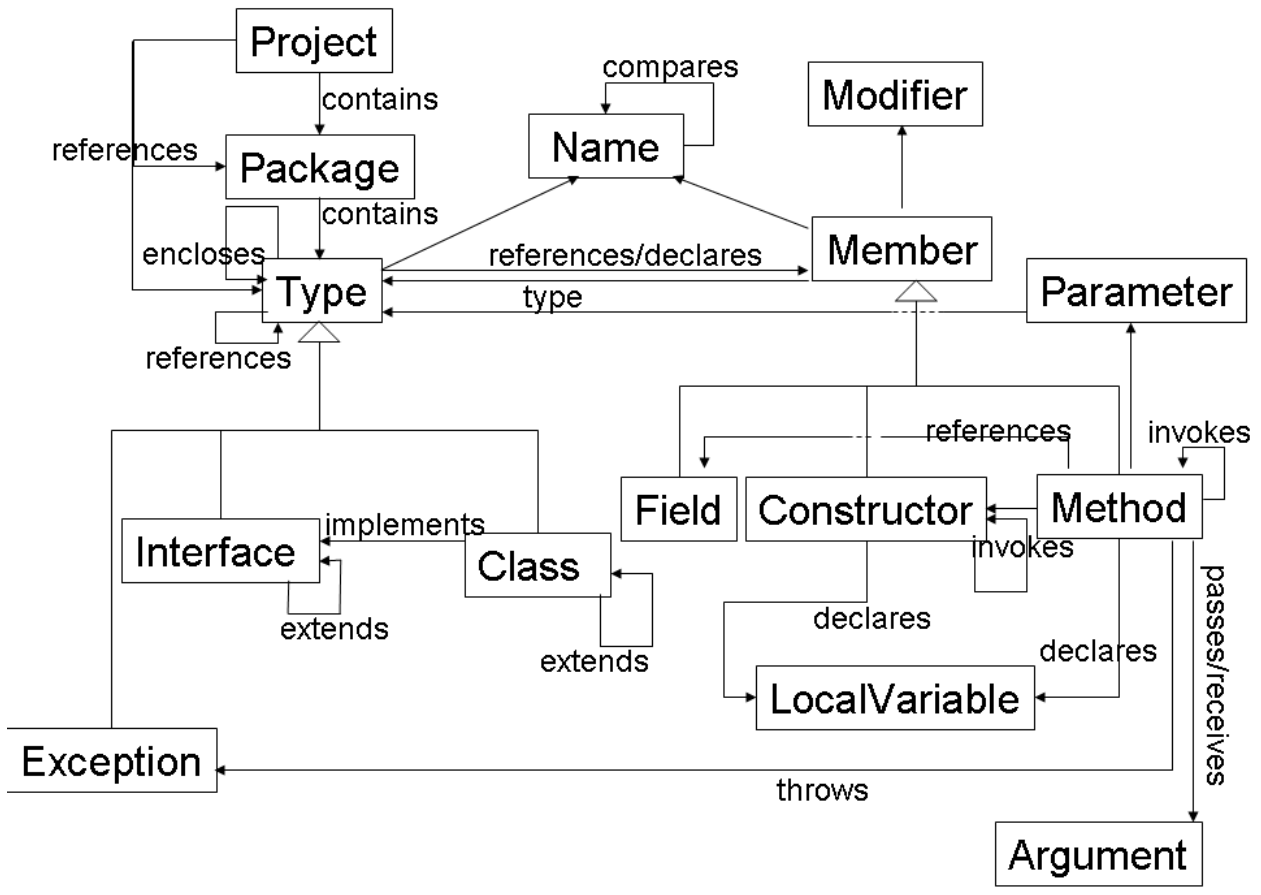

**Figure 3: Relationships relevant to sort queries**

The next section introduces the query component for describing relations specific to crosscutting concern sorts. It is up to the software engineer to organize these instances in more complex concern models that reflect her or his design decisions, and to provide appropriate identifier names.

# 4. DESCRIBING CROSSCUTTING CONCERN SORTS

## 4.1 The query model

The proposed query model is aimed at providing a standard, formalized description of the underlying relations of the crosscutting concern sorts. The model consists of a generic query definition to describe the query model, and a set of query templates (*sort queries*) to describe the relations specific to each of the sorts.

A sort query is a binary relation between two sets of elements, the *source* and the *target context*. A context is a (restricted) set of program elements, which can also be expressed as the result of a query. The two contexts in the definition of a sort query represent the *end-points* of the sort's relation.

The generic query describing the model can be expressed as:

```
relation_id (<source-ctx>, <target-ctx>);
```

The end-points contexts can be selected by the *sourceof* and *targetof* operators, respectively. Elements in the set of results of a query sort are (e1,e2) tuples, where:

```
(e1 in source-ctx) && (e2 in target-ctx) &&
    relation_id(e1,e2)
```

The two elements, e1 and e2, are program elements, such as a class or a method. The relation between them, *relation_id*, is a (binary) relation between program elements, like *call* or *inheritance* relations, which can be extracted by a static source code analysis.

The relevant relations for sort queries are shown in Figure 3. The same relations describe the list of predicates used by the sort queries. The *type* relation, between a Parameter or Member and a Type, shows the type of a field member or parameter, or the return type of a method. Note that *declares* relations can also be specified by using wildcards and qualified names, e.g., p.C.* for all members declared by class $C$ of package $p$.

In addition to these predicates, a query definition also allows for the transitive closure operator ($+$) and two "wildcards":

- * matches names of identifiers (element names or return types) or modifiers (visibility);
- .. matches any parameters or arguments in a method or constructor.

Variables are allowed and used to save (partial) results of a query, such as the set of elements that gives the end-point of a query. The variable can further serve as an end-point for another query. The notation used to declare a variable is <*variable_name*>.

The intersection of variables and end-point sets is shown through the && operator.

## 4.2 Template for the sort catalog entries

We describe the crosscutting concern sorts and their query component by an extended template that comprises the following elements:

**Intent** of the sort, to give a generic description of the sort's concerns;

**Relation** underlined by the sort, to describe the type of relation between the elements implementing instances of the sort;

**Idiom** specific to the sort implementation in an object-oriented (particularly Java) language;

**Query** associated with the sort to make explicit the relation between the elements implementing a sort's instance and to formalize the sort's definition;

**Example** to show a concrete instance of the sort;

**Other instances** to provide additional examples of (possible) occurrences of the sort.

## 4.3 Examples of sort instances

For exemplification of sorts we refer to a number of concerns identified in well known applications or libraries. Where examples of sorts instances are available, we refer to the JHotDraw application, which we have used as a case-study for documenting crosscutting concern instances using sort queries.

JHotDraw is a drawing editor for bi-dimensional graphics, developed as an open source project [3] with involvement from the authors of the "Design Pattern - Elements of Reusable Object-Oriented Software" book [8]. One of the application's goal is to provide a show case for good use of design patterns in the development of graphical applications.

## 4.4 The catalog

This section presents a catalog of crosscutting concern sorts described by the template introduced in Section 4.2.

### 4.4.1 *Method Consistent Behavior*

[3] jhotdraw.org

## Intent

Execute consistently a specific action for a number of method elements as part of their required functionality.

## Relation

The method elements share a (secondary) concern, which they implement through the consistent invocation of the action executing the desired functionality. The common action invocation shows the relation between the callers as participants in the implementation of a crosscutting concern.

## Idiom

Scattered method calls to the method implementing the common action to be executed consistently.

## Query

The sort query takes two arguments: a seed-element to define the source context, and the method implementing the action to be executed consistently by the elements in the source context. This method is the unique element of the target context.

```
<context> = pckg.I+ || (project Proj) ||
            type (pckg.Cls) || packge pckg;
```

We first define the (source) context based on the seed-element passed to the query, and save the elements of the context into a variable. The context definition shows several possibilites to use structural relations for describing the elements in the context; for example, the context could consist of the elements in the hierarchy of the type I of package *pckg* (pckg.I+), or the elements of the project *Proj*, or members of the type *Cls* of package *pckg*, or elements declared in package *pckg*, etc. The context seed is specific to each of these cases, and that is, the type I, the project *Proj*, the type *Cls*, and the package *pckg*, respectively.

The formalization of the context is similar to defining pointcuts in languages like AspectJ [13]. However, the context consists of program elements rather than execution points [4].

```
<selcallers> = <context> && sourceof(
                   invokes(method *, * p.C.m(..)));
CB(contextElem, m) = invokes(<selcallers>, * p.C.m(..));
```

Next, we save in the *<setcallers>*variable all the method members in the source context that invoke elements in the target context (in this case, the method *m*).

The query returns the invocation relationships between the elements in the source context and the method-action in the target context.

## Example

The notification mechanism in the previous example of the Observer pattern instance is an example of consistent behavior: the actions that change the state of the Subject object have to consistently call the notification method for allowing the observers to update their state accordingly.

A different example of this sort's instance is common to transaction management, a complex concern aimed at ensuring data integrity; this implies that an operation is committed only if it is fully completed and roll-backed otherwise. In a banking transfer operation, for example, both the debit and credit operation have to be successful at the same time for keeping a consistent state of the data. Java provides various alternatives to transaction management, like JDBC transactions and Java Transaction API (JTA) [5]. A JTA transaction, for example, implies that methods implementing the transaction logic consistently invoke dedicated methods of the *javax.transaction.UserTransaction* interface: the `begin` method at the beginning and the `commit` (or `rollback`) at the end to demarcate a JTA transaction. These invocations represent instances of the consistent behavior sort covering methods whose operations have to be under transaction management.

## Other instances

Logging of exception throwing events in a system; Wrapping of service level exceptions of business services into application level exceptions[17]; Credentials checking as part of the authorization mechanism[16]; etc.

### 4.4.2 Contract Enforcement

## Intent

Comply with design by contract rules, such as executing consistent condition checks to ensure that the assumptions about a method are confirmed.

## Relation

A set of elements are related by common condition checks.

## Idiom

Similar to Consistent behavior, instances of this sort are implemented as scattered calls to methods checking conditions as part of contract enforcement.

## Query

The sort is similar to the Consistent behavior sort with the difference that the action (condition check) is not part of the caller-elements' functionality; that is, a method can fulfill all its requirements in the absence of the calls checking the conditions. Hence, the two sorts differ only in intent. The query documents this intent by its own description while the same query template as for Consistent behavior applies to this sort too.

## Example

The use of assertions and assertions-like calls is a typical example of instances of this sort. An example of Contract enforcement instance is available through the JHotDraw drawing editor application. The application defines a hierarchy for Command operations to be executed as response to user actions, like menu items selection. The Command elements are participants in the implementation of the Command pattern, and define a no-argument `execute` method to carry out the specific action. Commands can be executed only if a valid view is present in the drawing editor. The Command elements check this condition through a consistent (check) call before proceeding with their execution.

## Other instances

Pre- and post-conditions complying with design by contract rules.

[4] The context definition shown for this query is common to several sorts to be discussed next. We will refer to the *<context>*variable, without redefining it.

[5] `http://java.sun.com/j2ee/1.4/docs/tutorial/doc/`

### 4.4.3 Entangled roles (Interfacing layer)

## Intent

Extend an element (method or type) with a secondary role or responsibility, which is entangled with its primary concern.

## Relation

The crosscuttingness of the concerns of this sort resides in the tight coupling between elements implementing different roles, like, for instance, relations between graphical user interface (GUI) elements and their associated model or action-controller elements. The GUI element knows its associated element and sends messages to it in order to be able to provide a response to an action or report about its state. The interfacing task of the GUI element results in its methods being crosscut by the logic of the model element. The methods' implementations address a different concern than the interface declaring them. The graphical elements store references to their model and mirror its state.

The relation specific to this sort is established between elements in the context of a (interfacing) class that invoke actions in the interfaced type through an object reference. This reference is typically stored in a field of the interfacing class.

## Idiom

Redirection of calls to a specific reference.

## Query

```
 //the type of the interfaced object
<interfacedType> = sourceof (declares(*, field p.C.field));
ER(C,field)=references(class p.C, <interfacedType>));
```

## Example

A common design practice for graphical user interface (GUI) elements, like Java Swing components, is to use a Model-View-Controller(MVC) design where the View and the Controller are implemented by the same class. The class for the graphical component keeps a reference to the model object, which determines the component's state. Hence, the graphical component is a wrapper around the model's state that "mirrors" the model's state and delegates actions to this model.

Swing components like buttons and menu items implement the described behavior (e.g., *AbstractButton*). The methods for setting or reporting about their state, such as *selected armed*, or *enabled*, redirect their calls to the model whose state they mirror.

The controller functionality in the same components delegates calls to its associated command-action. The action can also play a model role: in the JHotDraw drawing editor, a menu item is enabled or disabled if the Command object associated with the item is or not executable.

## Other instances

Swing GUI elements (like classes extending *AbstractButton*); Invoker-Command relation in the Command pattern.

### 4.4.4 Redirection layer

## Intent

Define an interfacing layer to an object (add functionality or change the context) and forward the calls to dedicated methods of the object.

## Relation

The redirection layer acts as a front-end interface that assumes calls and redirects them to dedicated methods of a specific reference, (with or without) executing additional functionality. The consistent (yet, method specific) redirection logic crosscuts this layer's methods.

Unlike the previous sort, instances of this sort do not query a model for its state, but typically add functionality dynamically to the calls. This sort is a specialization of the Interfacing layer sort.

The relation specific to this sort is between the redirecting layer and the target object, and resides in consistent redirection of calls between pair methods.

## Idiom

Identical logic in a number of methods that consistently redirect their calls to dedicated methods of a specific type.

## Query

```
//C-component; D-decorator
<compType> = sourceof(declares(*, field p.D.field));
 //all methods of C
<compMehtods> = targetof(declares(<compType>, method *));
//all decorator (D) methods
<decorMehtods> = targetof(declares(type D, method *));
RL(D,field) = (invokes(<decorMehods>,<compMethods>)) &&
              (compares(<decorMehods>,<compMethods>));
```

The relation is fairly complex and connects the context of the redirection layer and that of the type of the target-object.

The query takes as arguments the type element of the redirecting layer, and the field in this type that stores the reference to the object to redirect to. The two arguments define the source context, as the redirecting class, and the target context, as the type of the reference receiving the redirected calls.

The query returns invocation relations between methods with the same name in the two contexts. Most relevant for this relation are the methods in the redirecting layer that consistently forward their calls.

## Example

Implementations of the *Decorator* pattern are typical examples of instances of this sort. Decorator instances in JHotDraw allow to attach elements like borders to *Figure* objects. The decorators for figures extend *DecoratorFigure* that defines the set of methods to consistently forward their calls to the stored reference of the decorated object. Subclasses of *DecoratorFigure* can override its methods to dynamically extend its functionality.

## Other instances

Implementations of the *Decorator*, *Adapter*, and *Facade* patterns[17].

### 4.4.5 Add variability

## Intent

Use method-objects to pass a method as an argument.

## Relation

Instances of the sort implement a consistent mechanism of building and passing method-objects as method arguments. Method-objects are (typically) objects of a type declaring one method. The methods expecting arguments of this type only need and invoke the specific method for the passed object.

This mechanism is a substitute for passing pointers to methods as arguments. Languages like Java use this mechanism, which is also referred as *closures* or *functors* or *function objects*, to achieve a behavior similar to the use of call-back funtions.

The sort and its query describe a contract between client-callers and server-callees that make use of method objects as a substitute to passing references to methods.

## Idiom

A number of actions in a system require arguments of a method-object type for invoking the type's unique method. The clients conforming to the expectations consistently build and pass wrapper objects for the expected method.

## Query

```
//methods expecting args of MethodObject Type
<target> = sourceof(params(method *, type MethObjT));
<mCreateMethObj> = sourceof(invokes(method *, MethObjT.new(..)));
<source> = <mCreateMethObj> &&
           sourceof(invokes(method *, <target>));
//method-variables of MethodObject Type
<varsInMethods> = targetof(declares(<source>, var MethObjT *));
//methods being called with the previous variables as args
<target> = sourceof(args(method *, <varsInMethods>));
AV(MethObjT) = invokes(<source>,<target>);
```

The query documents the set of methods that create method-objects of a specified type, and pass them as arguments to methods declaring parameters of the same type. The target end-point of the relation consists of methods declaring parameters of a method-object type. The source context is given by the intersection between (1)the set of methods creating objects of a method-object type and (2) the set of callers of the methods in the target context.

## Example

Simple commands in JHoDraw, which do not save state, implement a number of operations like opening, printing and closing a drawing (application). The implementation follows the Command pattern and consists of associating menu items with instances of anonymous command classes that declare and implement an `execute` method to carry out the operation. These instances are method-objects and their `execute` method simply invokes a dedicated method that carries out the open (or close or etc.) operation. The Command object is hence used as a wrapper for the `execute` method to be invoked in response to menu item selection actions.

Other example instances are related to practices specific to functional programming [12], like the use of *closures* (*functors*) that can be implemented in Java using method objects like, for instance, *Runnable* objects. These objects only implement a `run` method and are used by common Java mechanisms, like thread safety. Consider, for example, *Component* elements (like Swing objects) that need to execute in a specific thread, i.e. the *event dispatching thread*, to avoid deadlocks during painting the graphical components. Two Java dedicated methods, `invokeLater` and `invokeAndWait`, ensure that these components execute in the special thread. The two methods expect an argument of type `Runnable` whose `run` method contains the code accessing functionality of the graphical (Swing) component to be executed.

## Other instances

Similar behavior as for the Swing components is present in the implementation and use of graphical elements of the Eclipse Standard Widget Toolkit(SWT). The *IRunnableWithProgress* interface, for instance, has to be implemented by classes with long-running operations for displaying a progress bar. The interface defines a `run` method, and instances of the implementing classes are passed as arguments to a dedicated method of a specialized class for running long-running operations.

SWT also defines an *user-interface thread* from which the SWT API methods to be called. Runnable objects are passed to this thread for (a)synchronous execution.

The Visitor pattern declares a specific *Visitable* role with an `accept` method to be invoked by visitor methods receiving arguments of type *Visitable*. Implementations of the pattern can rely on a (*Visitable*) method object to pass the `accept` method to interested clients.

Laddad proposes the *worker object* pattern and an implementation in AspectJ of the pattern [16]. The worker object is a method-object: an instance of a class implementing an worker method. Examples include Asynchronous method execution, Authorization using Java Authentication and Authorization Service (JAAS) API, etc.

### 4.4.6 Expose context

## Intent

Expose the caller's context to a callee by passing information to each method in the call stack of that callee.

## Relation

Instances of this sort are implemented by methods that are part of a call chain where an additional parameter is declared and used to pass a context along the chain. The declaration of additional parameters to pass context is crosscutting.

The elements related by this sort are the caller-method receiving an argument with a specific name (and type) and the callee-methods to which the caller passes the argument. To get a full description of the context passing, a transitive closure operator has to be applied to the relation associated with this sort.

## Idiom

Methods participating in a call chain declare additional parameters to pass a specific context required for fulfilling their (secondary) requirements.

## Query

```
<callees> = targetof(invokes(method p.C.caller(..), method *));
<selCallees> = <callees> &&
            sourceof(args(method *, name argName))) &&
            sourceof(args(method *, type ArgType)));
EC(caller, argName) = invokes(method p.C.caller(..),<selCallees>);
```

The query returns the relations between the input `caller` method and the callees that are passed an argument with the name specified in the query.

## Example

To monitor progress evolution for long-running operations in Eclipse applications, one can use a *IProgressMonitor* object. The long-running operations are passed a reference to the monitor class through an additional parameter. The operation invokes methods of this reference to indicate progress, like the `worked(int)` method to indicate that a given number of work units of the executing task have been completed. Long-running sub-operations receive the same reference to the monitor and use it to report progress.

## Other instances

Transaction management, Authorization, the Wormhole pattern [16];

### 4.4.7 *Role superimposition*

## Intent

Implement a specific secondary role or responsibility.

## Relation

A number of elements share a common role, other than their defining hierarchy. The sort is specific to classes that participate in multiple collaborations and hence implement multiple roles [21]. The concrete instances occur as multiple interfaces (or methods that can be abstracted into an interface) implementations. The crosscutting element resides in the tangling of multiple roles in the class implementing them.

## Idiom

Implementation (and definition) of members separable in distinct interface definitions; common instances occur as classes that implement multiple interfaces.

## Query

```
//role = class or interface
<implementors> = sourceof(implements (*, type Role));
<selectedImpls> = <context> && <implementors>
RSI(Role,contextElem) = implements(<selectedImpls>, type Role);
```

Role superimposition documents elements that implement or extend the interface or class defining a specific role.

## Example

Examples of role superimposition are available through implementations of design patterns defining specific roles, like the Observer pattern previously discussed, or the Visitor pattern.

Implementing persistence is also possible through role superimposition: the *Figure* elements in a drawing application, like JHotDraw, implement a *Storable* interface that defines the methods for a (figure) object to write and read itself to and from a file. Each figure implements these methods in a specific way to provide persistence and recovery of drawings over work session.

## Other instances

Implementations of design patterns that define specific roles; Implementations of multiple interfaces with dedicated, specific roles, e.g., *Serializable*, *Cloneable*, etc.

### 4.4.8 *Support classes for role superimposition*

## Intent

Make the relationship between classes explicit (through nested classes) to superimpose a role (to a hierarchy).

## Relation

A number of elements share a common role by enclosing support classes of a specific type.

Complex roles can be implemented through nested, support classes. The nesting mechanism enforces and makes explicit the relationship between the role of the enclosing class and the one implemented by the support class.

Instances of this sort occur as an alternative to multiple roles implementations: two hierarchies can interact by having common classes (classes that implement elements from both hierarchies) or by having elements from one hierarchy as support classes for the elements in the second hierarchy.

## Idiom

Hierarchies interaction through containment of nested classes.

## Query

```
<implementors> = sourceof(
            implements(class p.EnclosingC+.*, type Role));
SC(EnclosingC, Role) = implements(<implementors>, type Role);
```

The query associated with this sort reports about nested classes within a given context that implement a specific role. A common context is a class hierarchy to which the role is superimposed through the nested classes.

## Example

In the JHotDraw application there are two hierarchies that interact through support classes: the members of one hierarchy enclose members of the second, supporting hierarchy. The main hierarchy, *Command*, defines command elements for executing various application-specific activities like, copy and paste, or operations for setting the attributes of a figure. The second hierarchy, *Undoable*, defines operations for undo-ing and redo-ing the results of executing a command. Typically, each Command class encloses its associated Undo class.

## Other instances

Specialized iterators for various Collection types.

### 4.4.9 *Policy enforcement*

## Intent

Impose a (restriction) policy between groups of elements in the system.

## Relation

The sort describes a restrictive relation that limits certain type of interactions between a source and a target set of elements.

Concerns of this sort implement relationship policies between sets of elements that can be described in a source and target context, respectively. These policies cannot be enforced by language mechanisms, like, for instance, visibility. They are crosscutting because they have to be consistently documented and followed.

## Idiom

Requirements specifications available through documentation, source code comments, etc.

## Query

```
PE(srcContextElem, targetContextElem) =
            references(<src_context>, <tgt_context>);
```

The query documents (reference) relations between elements in a source and target context, respectively. If the enforced policy prohibits relations between the two contexts, the query will check the policy and report those elements that break it. This control query can be used as a complement to testing components.

## Example

Sun provides a list of restrictions imposed to enterprise beans[6]; one of these restrictions states that: "Enterprise beans should not use the *java.awt* package to create a user interface ... and stop the Java virtual machine and ...".

## Other instances

Typical examples of instances of this sort include interaction restrictions for (formalized) sets of program elements, such as packages, and classes implementing specific functionality. Eichberg et al. also discuss several examples of enforced policies [6]. Some of their examples, however, could be addressed by conditions checks, such as methods returning non-null values.

### 4.4.10 Exception propagation

## Intent

Propagate an exception for which no appropriate response is available.

## Relation

Concerns of this sort have specific a consistent propagation (re-throwing) mechanism of the checked exceptions thrown by methods that do not have appropriate answers to these exceptions.

Similar to context passing, the described relation is part of a call chain. The caller-elements implement the consistent (enforced) logic of re-throwing exceptions if not able to handle them. The transitive closure operator applied to this relation provides a full description of the elements in the call chain that re-throw exceptions.

## Idiom

Declare *throws* clauses in the definition of a method for passing the responsibility of catching the exception to the callers.

## Query

```
<callers> = sourceof(invokes(*, p.C.m(..)));
<throw> = sourceof(throws(method *, type p.ExceptionType));
<source> = <throw> && <callers> && <context>;
EP(m,ExceptionType,contextElem)=throws(<source>, p.ExceptionType);
```

The query reports relations between methods in a given context and the thrown exception of the specified type.

## Example

The file reading (writing) operations in JHotDraw implied by the drawing persistence concern, like `read(StorableInput)`, throw exception of the *java.io.IOException* type if the reading or parsing operation fails. The callers of these methods re-throw the same exception to their callers, up to the driver application that catches the exception and prints an error message.

## Other instances

Checked exceptions in Java

### 4.4.11 Design enforcement

## Intent

Enforce design, such as classes in an hierarchy must declare no-arguments constructors.

## Relation

A number of elements share a design logic (e.g., no-args constructors).

The sort describes concerns ensuring design rules compliance for program elements in a defined (formalized) context. The implementation of these concerns lacks language support, and relies on documentation, specifications, comments, etc.

## Idiom

Requirements specifications available through documentation, source code comments, etc.

## Query

```
//method
DE(contextElem,m)=declares(<context>, member * m(..));
```

The design enforcement documents a relation between a set of elements that (have to) declare specific members with various signatures.

If member $m$ is a constructor, the *new* keyword will be used instead of the method name. The member can be a field as well.

## Example

The documentation for the *Storable* interface in JHotDraw to be implemented by elements that can be stored and restored to and from a file states that "objects that implement the interface ... have to provide a default constructor with no arguments". Similar to JavaBeans design, the no-argument constructor will be invoked to create objects after being read from the file.

## Other instances

Serialization rules in Java require that (special) serializable classes implement special methods (`writeObject(..)` and `readObject(..)`) with specific signatures. These methods are not declared by the *Serializable* interfaces but the design requirements are provided through the documentation of the interface.

JavaBeans design (providing classes with no-argument constructors) is similar to the discussed persistence example.

### 4.4.12 Dynamic behavior enforcement

[6] `java.sun.com/blueprints/qanda/ejb_tier/restrictions.html`

### Intent

Enforce rules for object use, like before-use initialization and post-use clean-up.

### Relation

Shared logic of manipulating objects of a certain type.

Instances of the sort implement rules for object use, like the use of lifecycle methods.

### Idiom

Requirements specifications available through documentation, source code comments, etc.

### Query

```
<context> = targetof(contains(type C, method *));
DBE(C,field)=set(<context>, C.field) || get(<context>, C.field);
```

We assume a (crosscutting) solution aimed at checking that certain methods of an object are executed in an expected order. The solution consists of declaring a field which to be set to a different (integer) value by each executing method of the field's object. The value of the field cannot be lower with more than one unit than the value to which a method is going to set the field. That is, each method checks the value of the counter field before executing in order to ensure proper order of the execution.

### Example

(De-)Activation of tools in the JHotDraw graphical editor requires that the `deactivate` method is always used to clean-up resources when a tool has finished its execution and the user switches to a different tool. This method should be the last called for a given tool.

### Other instances

Lifecycle concern [17];

## 5. COVERAGE

This section investigates how the sorts are able to cover the crosscutting concerns present in complex structures, commonly referred in literature for their crosscutting properties. To this end we consider the list of design patterns for which Hannemann and Kiczales reported improved modularity due to proposed aspect-oriented (AspectJ [1]) solutions [10].

Table 1 shows the list of these design patterns. Each pattern corresponds to a composite concern model whose crosscutting elements are described by sort instances relations. The documented crosscutting elements in the design patterns have been identified by examining the description of the patterns and the sample code in Gamma et al. [8], as well as the aforementioned AspectJ solutions [7]

### 5.1 Design patterns

Implementations of the *Adapter* pattern could use either multiple roles or object composition to adapt a class to an interface expected by clients. In the first case, the *Adaptee* role is super-imposed to the class implementing the Adapter functionality. The Adapter class implements both a Target interface and (extends) the Adaptee, which is an instance of the Role superimposition sort.

The solution relying on object composition would typically use delegation from the Adapter to a stored reference of the Adaptee object. This is an instance of the Redirection layer sort.

The *State* pattern comprises a number of crosscutting elements: The *Context* role is super-imposed and has specific members for maintaining a reference to the object defining the current state; second, the notification of changes of the current state to be stored in the Context object is an instance of the Consistent behavior sort. The third element is an instance of the redirection layer: the Context object forwards the received calls to the methods of the object storing the current state.

The crosscuttigness occurring in the implementation of the *Decorator* pattern can be described by the Redirection layer sort. The methods in the decorator class consistently redirect their calls to dedicated methods in the decorated class, through the stored reference to the decorated object. Decorators are typical examples of instances of the Redirection layer sort.

The crosscutting element of the *Proxy* pattern implementation resides in the consistent forwarding of the calls to the reference of the real subject class, stored by the Proxy object.

Another crrosscutting concern occurs in *protection proxies* as an instance of Consistent behavior; this consists of a method call that checks the access permissions before executing the forwarding operation. Some implementations also consistently check if the proxy's subject has been initialized. This check is part of the method for accessing the reference to the subject. The method is invoked by the actions in the proxy that forward their calls.

The *Visitor* and *Composite* patterns are often used in combination [8, 10]. Both patterns define roles that in various implementations are super-imposed like, for instance, the *Visitable* role. The roles we chose to document by sort instances, *Composite* and *Visitable*, are those defining role-specific methods.

Certain implementations could make use of method objects for passing a reference of the *Visitable* object and its `accept` method to the visitors' methods. Such an implementation could be documented by an instance of the Add varaibility sort. We show in Appendix A an AspectJ solution for the Visitor pattern, different from the one of Hannemann and Kiczales [10], that addresses the Add variability concern.

The *Command* pattern presents several participants, like the Command ojects, the Invokers that require the execution of the command, and the Receivers to carry out the requests. These participants play specific roles in the implementation of the pattern; however, not all of them declare specific members and end up as super-imposed.

The Invoker participant keeps a reference to an associated command to which it forwards the requests. Invokers like buttons and menu items in graphical user interfaces are often interfacing the command object by mirroring its state through their own display (e.g., enabled buttons correspond to commands that can be executed with the active configuration, etc.).

In some implementations, the Invoker role is super-imposed through interfaces that declare methods dedicated to handling events that trigger execution of commands. The implementations of these methods in the command invokers consistently call the operation in the Command class. The behavior indicates an instance of the Consistent behavior sort.

[7]The solutions are available as a set of simple examples of design patterns implementations in Java and AspectJ at `www.cs.ubc.ca/~jan/AODPs/`. The crosscutting concerns discussed in this paper are, however, not limited to these examples only.

| Design pattern | Composition of sort instances |
|---|---|
| Adapter | Adapter = RSI(Adaptee, contextElem) + RL(Adapter, adapteeReference); |
| State | State = RSI(Context, contextElem1) +<br>CB(contextElemStateChanger, Context.changeState(State)) +<br>RL(Context, stateReference)); |
| Decorator | Decorator = RL(Decorator, componentReference); |
| Proxy | Proxy = RL(Proxy, fieldRefRealSubject);<br>*Protection proxies:*<br>*document the consistent behavior of checking the credentials:*<br>CB(contextElem, checkAccessPermission()); |
| Visitor | Visitor = RSI(VisitableElement, contextElem);<br>*Specific implementations:*<br>Visitor = AV(VisitableElement); |
| Command | Command = RSI(Receiver, contextElem1) +<br>ER(Invoker, commandReference) +<br>RSI(Invoker, contextElem2) +<br>CB(invokersContext, Command.execute()));<br>*Certain implementations using Command for method objects:*<br>AV(Command); |
| Composite | Composite = RSI(Composite, contextElem);<br>*RSI(Leaf, contextElem2)) - not crosscutting* |
| Iterator | Iterator = RSI(Aggregate, contextElem); |
| Flyweight | Flyweight = RSI(Flyweight, contextElem1) +<br>CB(contextElem2, FlyweightFactory.getFlyweight)); |
| Memento | Memento = RSI(Originator, contextElem1) +<br>CB(careTakerContextElem1, Originator.createMemento)); |
| Strategy | Strategy = RSI(Context, contextElem);<br>*sometimes, we could also have:*<br>RSI(Strategy, contextElem1); |
| Mediator | Mediator = RSI(Colleague, contextElem) +<br>CB(contextElem, notifyMediator)); |
| Chain of Responsibility | ChainOfResponsibility = RSI(Handler, contextElem1) +<br>RL(Handler, successorReference)); |
| Prototype | Prototype = RSI(Prototype, contextElem);<br>*In some languages, like C++, copy constructors are required:*<br>DE(contextCloneableObjs, CloneableType.new(const CloneableType&));<br>*A similar instance can be used for requiring implementation*<br>*of the Object.clone method in Java* |
| Singleton | Singleton = RSI(Singleton, contextElem1) +<br>DE(contextElemSingleton, private Singleton.new(..)) +<br>CB(contextElem2, Singleton.instance()); |
| Observer | Observer = RSI(Observer, contextElem1) + RSI(Subject, contextElem2) +<br>CB(notify, contextElem3)+<br>CB(attachObserver, contextElem1)+ CB(dettachObserver, contextElem1); |

**Table 1: Design patterns as compoostion of sort-instances.**

Particular implemenations of the Command pattern can serve as method-objects. In this case, the class defines just an *execute* method and does not save state. The Command object is passed around for clients to access the functionality of its unique method. The use as method object is an instance of the Add variability sort.

The only crosscutting element occurring in the implementation of the *Iterator* pattern, is the super-imposed *Aggregate* role. The role defines the `CreateIterator()` method to create an iterator object for traversing the elements of the aggregate (structure).

The concerns documented for the *Flyweight* pattern comprise a Role superimposition instance for the *Flyweight* role, and a Consistent behavior for obtaining references to a (new) flyweight object. This behavior consists of calling the accessor method in the factory class for the flyweight instances, instead of attempting to build new flyweight objects. A similar behavior is discussed below for the *Singleton* pattern.

The refactoring of *Memento* pattern to AspectJ uses the introduction mechanism for superimposing the *Originator* role. In addition to this, we document a Consistent behavior instance, namely acquiring a memento object before performing the operation that changes the state.

The *Strategy* pattern defines two roles, the Strategy and the (Strategy)Context. Most commonly, the Context is a super-imposed role, maintaining a reference to the Strategy object (and defining methods to access the reference) and, possibly, delegating requests from its clients to the Strategy.

The *Mediator* pattern implies a super-imposed role (*Colleague*) to store and access the reference to the *Mediator* class. Moreover, each change in the coleague class results in a consistent notification of the mediator for coordinating the other colleague-classes. In some implementations, the *Mediator* role could also be super-imposed.

The participants in the responsibility chain are superimposed the Handler role, which defines the method for handling specific requests, and the reference to the next Handler in the chain. The handler-methods check the request and consistenly pass it to the following handler in the chain. Because the one-to-one association between the passing and receiving methods in the chain, the call passing behavior is documented by a Redirection layer instance.

One crosscutting element in a typical implementation of the *Prototype* pattern is due to the super-imposition of the *Prototype* role that declares the *clone* method to allow objects to generate copies of themselves. In some languages, like C++, the Prototype must declare a copy constructor for cloning. (A copy constructor receives as unique parameter a constant reference to an object of the class' type.) This would be an instance of the design enforcement sort.

In Java, the cloning is realized through the *clone* method in the *Object* class, which is extended by all Java classes. The class implementing an overriding method of `clone` has to implement the *Cloneable* interface to indicate to the *clone* method that it is legal to make copies of the fields of the Cloneable class. This is a design enforcement sort instance.

The *Singleton* pattern has specific the access method to the unique instance of the singleton class, which has to be used by clients instead of calling a constructor. In some implementations, the method is declared by an interface that defines the super-imposed role of the Singleton.

Singletons have special designs, most notably they have to declare the constructor as private for not allowing constructor calls from outside the class. This Design enforcement instance cannot be specified other than by comments.

Similar to the Flyweight pattern, we use a Consistent behavior query to document the references to the method returning the unique instance. This behavior shows the rule for accessing the singleton's functionality.

The *Observer* pattern is documented as a composition of Consistent behavior and Role superimposition intances, as discussed in the previous sections. In addition to the consistent behavior of notifying changes in the Subject's state, we also document the mechanisms for registration and deregistration of observers.

## 6. TOOL SUPPORT

SoQueT (SOrts QUEry Tool) is the tool support we built to query for sort instances and organize related instances in composite concern models [8]. The concern model supported by SoQueT is similar to that in CME, with the differences that complex concerns are expressed as compositions of sort instances [11].

SoQueT provides three main user-interface components: The interface to define a query for a specific sort based on a list of pre-defined templates is shown at the bottom of Figure 4. The template guides the user in querying for elements that pertain to a concrete sort-instance.

The results of the query are displayed in the *Search Sorts* view, also shown in Figure 4. The view provides a number of options for navigating and investigating the results, like display and organization layouts, sorting and filtering options, source code inspection, etc.

The *Concern model* view allows to organize sort instances in composite concerns described by their user-defined names. The concern model is a connected graph, defining a view over the system that is complementary to the Eclipse's standard package explorer. The graph is displayed as a tree hierarchy, with sort instances as leaf elements. A sort instance element can be expanded to display the results of its associated query. The node tree representing a sort instance is labeled with a user defined name and the description of the associated query. Note that queries can be associated only to sort instances and not to a composite concern.

The tool introduces the concept of *virtual interface* to support documenting sorts like Role superimposition. Because some super-imposed roles do not have dedicated interfaces but are tangled within other interface or class declarations, the tool allows to define virtual interfaces for these roles. The user can select in a graphical interface those members of the multi-role type that define the analyzed role.

For some of the sorts that rely on a reference to a type for specifying relations, like Redirection or Interfacing layer, the referred type can be specified as either the type of a field or the return type of a method in the referring type.

## 7. SORTS IN PRACTICE

We have documented crosscutting concerns in the JHotDraw application to test the suitability of the sorts and their associated queries for describing and documenting such concerns in real systems.

JHotDraw is a relevant case for this task as it has been developed as a show-case for applying design patterns, and as we have seen, many of these patterns involve crosscutting functionality. Moreover, the application has been proposed and used as a common benchmark for aspect mining [17, 2].

We focus the discussion in this section on various instances of design patterns in JHotDraw and how they can be modeled in SoQueT. The model for the documented concerns in JHotDraw, including all those described in this paper, is available for download [9]. This model can be loaded into SoQueT and used to support concern understanding and exploration for the selected application, as well as a reference for aspect mining.

### 7.1 Design patterns in JHotDraw

Figure 5 shows a number of core elements in the JHotDraw application and their collaborations. The figure is aimed at supporting the understanding of the pattern implementations discussed next.

[8] `http://swerl.tudelft.nl/view/AMR/SoQueT`
[9] `http://swerl.tudelft.nl/view/AMR/SoQueT`

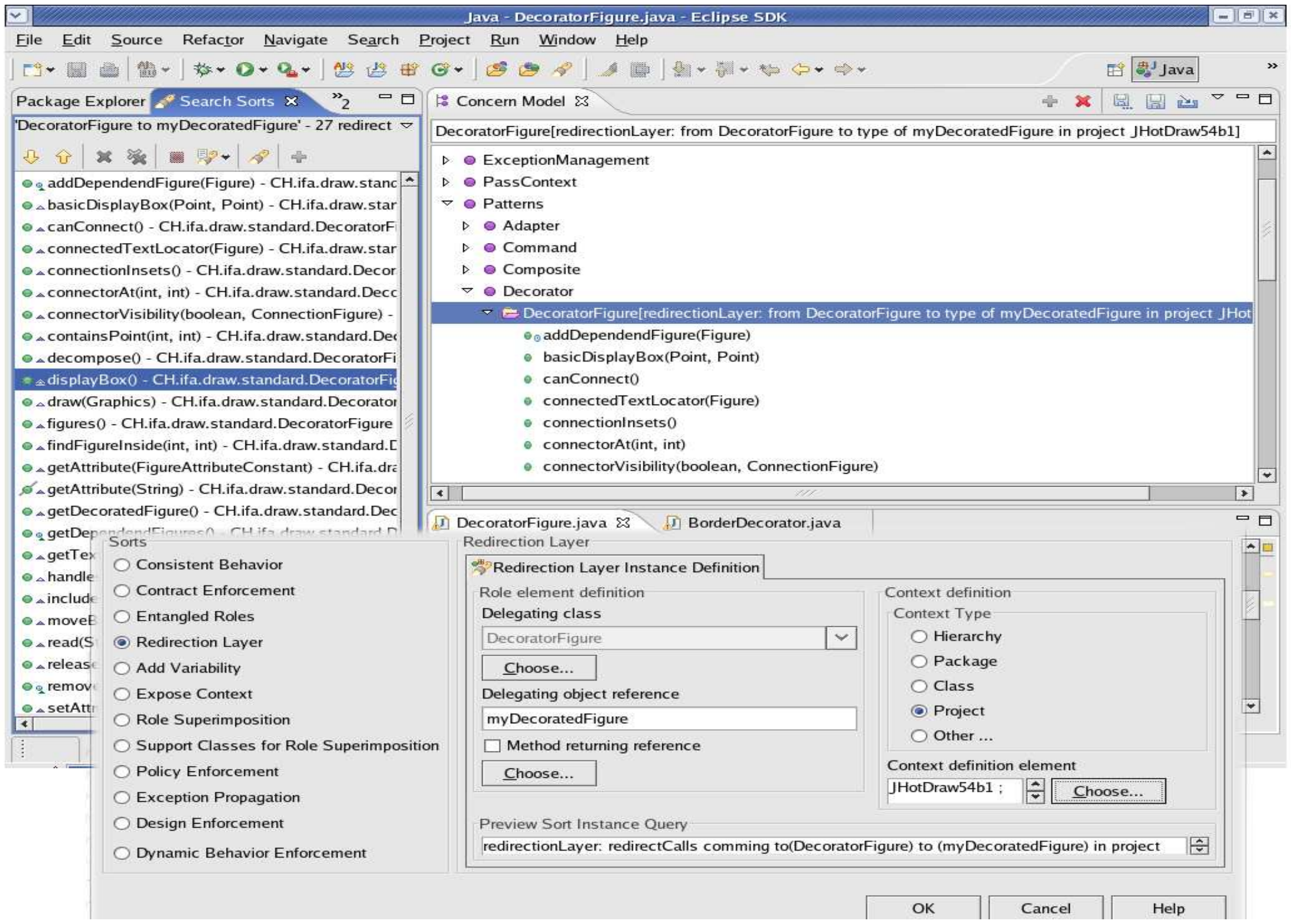

**Figure 4: SoQueT views**

## Strategy

The *Strategy* pattern in JHotDraw is implemented by a number of elements, like *Connector*s and *Locator*s. Connectors (e.g., *ChopEllipseConnector*, *PolyLineConnector*, etc.) define strategies for locating connection point on a figure, while locators (*ElbowTextLocator*, *PolyLineLocator*, etc.) locate a handle position on a figure.

The *Strategy* role is a primary role for the interfaces defining it and the classes implementing these interfaces. The *Context* strategy role, however, is super-imposed, through implementations of the *ConnectionFigure* and *LocatorHandle* interfaces respectively.

We document the pattern-specific roles by instances of the Role superimposition sort: a sort query asks for all implementors of *ConnectionFigure* to describe the Context, and a second query documents the Strategy in a similar way. We group these two instances in a composite concern bearing the name of the pattern instance, *ConnectionStrategy*.

Another instance of the pattern implements the update strategy for drawing views. The *Painter* interface defines the strategy in the context of the *DrawingView* hierarchy. The *DrawingView* interface, however, defines multiple roles and participates in multiple implementations of the *Strategy* pattern for which it defines the context role. To document the role of interest, we use a virtual interface definition. The virtual interface definition comprises the members that are role-specific, namely the members to refer (maintain and allow access to) the Strategy object.

## Visitor and Composite

The two patterns can be discussed together for the JHotDraw case, too. The *Visitor* pattern in JHotDraw allows to, for instance, insert or delete figures into or from composite figures, such as drawings. To realize this functionality, the top level interface for figure elements (*Figure*) defines the *Visitable* role. The role is tangled with the other concerns common to figures, such as drawing the figure.

The one member of *Figure* specific to the Visitable role is `visit(FigureVisitor)`[10]. This will be part of a virtual interface definition for the super-imposed role of Visitable for figure elements. Additionally, we can document the *Visitor* role defined by the *FigureVisitor* interface.

One of the two implementations of the method for accepting visitors corresponds to composite figures and passes the visitor to all the child-figures. Composite figures in JHotDraw extend the *CompositeFigure* abstract class, which defines and implements the method for manipulating child-figures. These methods are specific to the Composite role, in the pattern with the same name, but tangled with the definition of figure-specific methods. A new virtual interface definition for Role superimposition, which to include the methods for children manipulation, helps us to document this crosscutting concern.

The Add variability instance discussed for the Visitor pattern in Section 5 is not applicable to this implementation of the pattern.

[10] The name of the method could be misleading, as it corresponds to the *accept* method in the GoF desciption of the pattern

## Decorator

JHotDraw supports decoration of figures with borders and animations through the use of the Decorator pattern. An abstract class, *DecoratorFigure*, defines the decorator interface and the set of methods forwarding the requests to the decorated figure. Concrete decorators extend the interface and override some of the methods to add specific functionality.

The pattern's implementation is a standard example of Redirection layer sort instance. We document the instance accordingly, by specifying in the query template the *DecoratorFigure* type and its field that stores the reference to the decorated *Figure*. The query returns all the methods in the Decorator that consistently redirect their calls to the reference of the decorated figure.

The specification and the results of the query are also shown in Figure 4.

## State

The *Tool* elements in JHotDraw realize an implementation of the *State* pattern in the the context of *DrawingViews*. The *Tool* and *DrawingView* interfaces define the *State* and *Context* roles, respectively, which we describe by simple instances of the Role Superimposition sort.

The collaborations in the implementation of this pattern instance are, however, more complex: A *DrawingView* knows its *DrawingEditor* and through this it gets access to the active *Tool* object. The tools consistently notify the editor at the end of their interaction by invoking the editor's `toolDone` method. The notification results in re-setting the active tool to the default one. The consistent (notification) behavior is a concern that crosscuts the tool elements. To describe the concern using sorts, we query for the calls to the editor's `toolDone` method and restrict the query's context to the *Tool* hierarchy.

Most of the documented calls originate from mouse events in the *Tool* classes. The mouse events are initiated by the objects defining the State context, like a DrawingView forwards all its input events to the editor's active tool. The forwarding operation is an instance of the Redirection layer sort. To document this concern, we query for the references from mouse events listeners in the context class to the return type of the method providing the access point to the tool (i.e., `Tool DrawingView.tool()`). This query, however, will return no result because of the signature mismatch between the redirecting (delegator) method, and the method to redirect to: for example, the `mouseReleased(..)` method of the listener calls the tool's `mouseUp(..)` method, and `mousePressed` calls `mouseDown(..)`. This is a limitation of the tool in documenting redirection behavior.

## Mediator and Observer

The *DrawingEditor* element discussed above is participant in an implementation of the *Mediator* pattern by defining the interface to coordinate the different objects that participate in an editor. The participating elements (i.e., the *Colleague* objects) are *Tool*, *Command*, or *DrawingView* objects. Each of the three types defines accessor methods for the mediator object: these methods are part of the *Colleague* role and we document them as instances of the Role Superimposition sort by declaring virtual interfaces for the methods of interest.

The *Tool* and *Command* elements communicate with the *DrawingEditor* mediator by using the *Observer* pattern: the two types of elements register themselves as listeners of the mediator and receive notifications of changes of other colleagues. The *DrawingEditor* interface defines the *Subject* role in the context of the *Observer* pattern for allowing the mediator to communicate with the colleague objects. We document this role by using (a virtual interface definition for) Role Superimposition. This also documents the *Mediator* role. The calls to the notification mechanism (the `figureSelectionChanged` method in *DrawingEditor*) are an instance of the Consistent behavior sort. The context for the Consistent Behavior instance is given by the union of the three hierarchies defining the *Colleague* roles: *Tool*, *Command*, and *DrawingView*.

This is one of the several instances of the Observer pattern that we have documented, one of them being shown in Figure 1, for listeners to changes in figure elements.

## Prototype

One implementation of the *Prototype* pattern covers *Figure* elements and specific *Tool* elements, namely *CreationTools* to create new figures from a specified prototype. The *Figures* explicitly (re-)declare the inherited *Object*'s `clone` method; hence, documenting the *Prototype* role using sort instances only requires to define a virtual interface declaring this method. To allow cloning using the Java mechanism, the *Figure* interface has to extend Java's *Cloneable* interface. Although the interface does not declare any member, the extension declaration is part of a crosscutting concern that we document by a Role superimposition instance realized through the *Cloneable* interface in the context of the *Figure* hierarchy.

Other client types using (particular) *Figure* prototypes include *ConnectionTool* and *ConnectionHandle*.

## Command

The *Command* pattern is implemented by 40 elements in the *Command* hierarchy; half of these are anonymous classes. The *Command* classes implement the *Command* role in the pattern. The role can be described as an instance of the Role superimposition sort.

The *Receiver* functionality varies for the different commands: some of them have associated a specialized Receiver, while others directly implement the command's logic (e.g., *AlignCommand*). A common receiver for commands are the *Figure* or *Drawing* elements as, for instance, for *BringToFront* and *SendToBack-Commands*. However, many commands do not carry out only a single forwarding, and the actions they delegate to are not always dedicated to the delegating command; that is, describing the *Receiver* role as super-imposed is often a per-case decision.

We document the *Receiver* role, for instance, in the *DrawApplication* class. The class implements a method to create the standard menus of the JHotDraw drawings editor. Each menu is associated an anonymous command whose `execute` method delegates execution to dedicated actions in the *DrawApplication* class. These actions define the *Receiver* role, which we describe in a sort-query using a virtual interface definition (in the context of the same class).

A group of Command-*Invoker* elements consists of MenuItems and Buttons. These elements interface the associated commands by listening for action events and triggering the execution of commands. We document this behav-ior with instances of the Interface layer sort. Furthermore, the Command invokers implement the *ActionListener* interface whose unique method, `actionPerformed`, consistently invokes the `execute` method of the associated command. This is an instance of Consistent behavior.

The Command hierarchy also exhibits several other crosscutting concerns, less relevant to the pattern itself: The named commands conduct a pre-condition check before execution, and a consistent notification at the end of their execution. These concerns are instances of the Contract enforcement and Consistent behavior sorts, respectively. Commands make use of support classes to implement undo functionality, as discussed in Section 4. The undo support also requires a proper, consistent initialization, which is a consistent behavior crosscutting the `execute` methods of the *Command* classes.

## Adapter

The *Adapter* pattern is implemented by the *Handle* elements, which adapt *Figure*s to a common interface accessed by *Tool* objects. Handles allow tools, like selection tools, to manipulate figures. The implementation of the Adapter pattern is based on object composition: the handle element stores a reference to its owning figure and defines an accessor method to this reference. The tool client (*HandleTracker*) delegates mouse events to specific methods in the *Handle* interface, which in turn translates the events into actions directed to the handle's figure.

The identified crosscutting elements to be described by sort queries include the *Adapter* role in *Handle* classes to access the reference to the Adaptee (i.e., *Figure*). We use for this a virtual interface definition that declares the accessors method for the Adaptee reference.

Although the implementation of handles is aimed at translating events into actions for figure elements, it does not follow consistent rules for redirecting functionality to the adaptee. Such a rule is apparent in the implementation of the client tool: mouse events are redirected to dedicated methods declared by the *Handle* interface; however, each mouse event (up, drag, down) delegates to a method with a different signature (invokeStart, invokeStep and invokeEnd). Due to limitations in the tool implementation (and the query template for the redirection layer sort), the concern remains undocumented.

Nevertheless, *Handle*s act as an interfacing layer for figures: the actions carried out by handles are in fact operations in the *Figure* objects. For instance, the *PolygonScaleHandle* for scaling and rotating *PolygonFigure*s relies on the `scaleRotate` method in the *Figure* class to answer mouse-drag events.

## Iterator and Singleton

The *FigureEnumertor* class participates in the implementation of multiple patterns. The class acts as an *Adapter* for the standard Java *Iterator*, implements the Singleton pattern, and the *Aggregate* role in the Iterator pattern.

The *Singleton* role is documented through a virtual interface declaring the singleton member and the method to access it. We further document the references to this method as a consistent behavior for accessing the functionality of a *FigureEnumerator* object. A design enforcement sort instance indicates that the class should declare (only) a private constructor. However, this implementation of the pattern

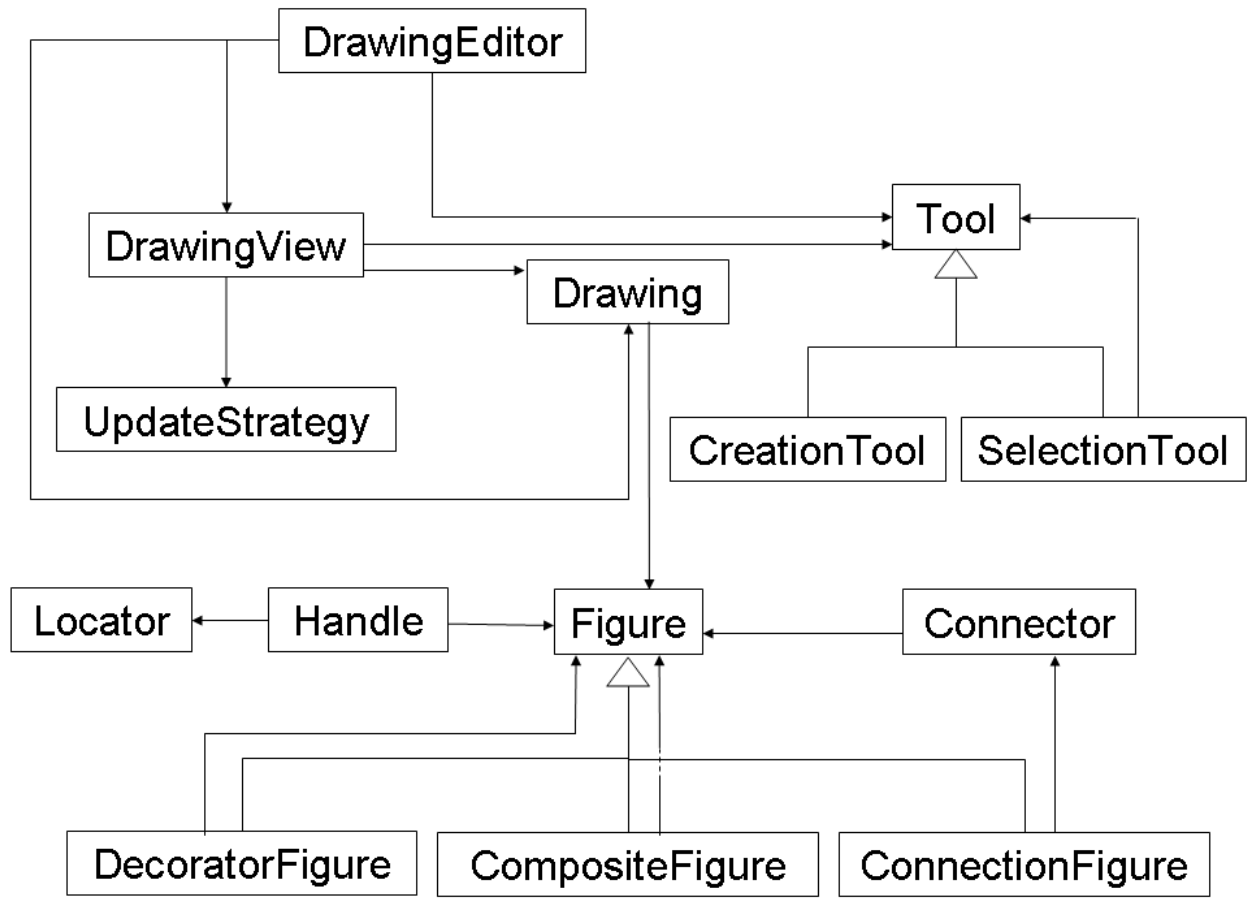

**Figure 5: Collaborations in JHotDraw**

uses a public constructor for resetting the Singleton. This is rather atypical.

The access method to the unique instance of *FigureEnumerator* is also part of the *Agregate* role in the context of the Iterator pattern, which is documented by a Role superimposition instance.

## 8. DISCUSSION

The experiments conducted for describing crosscutting concerns as sort instances in complex cases, like design patterns, proved that sorts were effective in most of the cases. The analysis carried out on different sample implementations of design patterns [8, 10], as well as on real systems, like JHotDraw, shows that sorts cover well the crossccuting concerns described in literature by various authors. Moreover, we covered a larger variety of concerns with the examples shown in the proposed catalog.

The sort queries and their tool support can be improved in several cases. The query for the Redirection layer sort, for instance, was not flexible enough for the cases where the methods in the redirecting layer had different names than the target methods. This limitation is due to the name matching criteria, also common in many aspect-oriented languages, such as AspectJ. However, this matching is a common practice and we expect it to be present in most cases.

The formalization of the context to restrict the domain of a relation can benefit from more flexible definitions as well. Although the tool support for sort queries allows to define contexts as enumerations of program elements, like classes, packages, etc., elegant formalization of contexts give a clearer description of the concern. Since context definitions are often similar to pointcuts definitions in aspect-oriented languages, improved support for defining pointcuts is equally useful for expressing contexts.

We mainly used structural relations for formalizing the definition of a query's context; however, many aspect (query) languages also allow for definitions that use name-based criteria, such as all methods whose name starts with *get*. Another extension could consist of using support for source code annotations to relate elements in a context by a common intent, like modifiers of the Subject state in implementations of the Observer pattern.

The catalog of crosscutting concern sorts can contribute to aspect refactoring efforts and extension of query languages for aspect-oriented programming. For instance, sorts like Design enforcement cannot be expressed by CME's query language or aspect languages like AspectJ.

The sort instances allow to group elements participating in relevant crosscutting relations, which are not explicit in source code. In this respect, the sorts are modular units comparable with aspects. However, sorts are mainly aimed at supporting crosscutting concern comprehension by describing atomic elements in a standard, consistent way. Such elements can be associated template refactoring solutions based on their description provided by sort queries. This is a first step towards refactoring concerns to more complex aspect solutions.

The sorts describe crosscutting concerns both intentionally and extentionally. The extent of the concern consits of the elements captured by the sort query. The intent is given by the query itself. New elements to be added to a system should be aware of the intent of the concerns documented by sort queries and be consistent with existent rules and policies. To this end, the user should be able to query the concern model for concerns that "touch" program elements of interest to a development or maintainance task.

## 9. CONCLUSIONS

This paper proposed a system for addressing crosscutting functionality in source code based on crosscutting concern sorts. Such a system can provide consistency and coherence for referring and describing crosscutting concerns. As a result, sorts are useful in program comprehension and areas like aspect mining and refactoring.

We have described crosscutting concern sorts as relations between sets of program elements and formalized these relations through a query component. We have organized the sorts in a catalog and discussed each sort in significant detail, describing specific implementation idioms and examples of concrete instances. Sorts have been assessed for crosscutting relations present in design patterns and in a real application system.

## 10. ACKNOWLEDGMENTS

I thank Arie van Deursen, Juri Memmert and anonymous reviewers for comments and feedback on this and earlier versions of this work.

# APPENDIX

## A. AN ASPECTJ SOLUTION TO THE VISITOR PATTERN

This example shows a simplified AspectJ solution for a pricing visitor: any call to the `getPrice` method of a *(Composite)Equipment* object are advised for building a *Visitable* object and pass it to visitors that just invoke the `accept` method of the passed argument.

```
pointcut price(Equipment equipment):
    call(int Equipment+.getPrice())
    && target(equipment);

int around(final Equipment equipment) : price(equipment) {
    Visitable visitable = new Visitable() {
        public void accept(EquipmentPriceVisitor v) {
            if(equipment instanceof CompositeEquipment) {
                List subcomponents = equipment.getComponents();
                for(int i=0; i<subcomponents.size(); i++)
                    (((Equipment)subcomponents.get(i))).getPrice();
            }
            v.addToPrice(proceed(equipment));
        }
    };
    visitor.visitEquipment(visitable);
    return visitor.getTotalPrice();
}
```